# Mutual control of critical temperature, residual resistance ratio, stress, and roughness for sputtered Nb films

E.V. Zikiy,[1,2] I.A. Stepanov,[1] S.V. Bukatin,[1] D.A. Baklykov,[1] M.I. Teleganov,[1] E.A. Krivko,[1] N.S. Smirnov,[1] I.A. Ryzhikov,[1,3] S.P. Bychkov,[1] S.A. Kotenkov,[1] N.D. Korshakov,[1] J.A. Agafonova[1] and I.A. Rodionov[1,2,*]

[1]Shukhov Labs, Quantum Park, Bauman Moscow State Technical University, Moscow, 105005, Russia

[2]Dukhov Automatics Research Institute (VNIIA), Moscow 127055, Russia
[3]Institute for Theoretical and Applied Electromagnetics RAS, Moscow 125412, Russia
*email: irodionov@bmstu.ru

**Abstract**

Superconducting single quantum logic integrated circuits traditionally exploit magnetron sputtered niobium thin films on silicon oxide substrates. The sputtering depends on multiple process parameters, which dramatically affect mechanical, electrical and cryogenic properties of Nb thin films. In this work, we focus on the comprehensive relationship study between 200-nm Nb film characteristics and their intrinsic stress. It is shown that there is a critical value of the working pressure $p_{critical}$ at the fixed sputtering power above which stress in the film relaxes whereas the film parameters decrease. Below $p_{critical}$ one can control intrinsic stress in the wide range from -400 MPa to +600 MPa maintaining low surface roughness (Rq) 0.8 nm, electrical resistivity less than 20 μΩ×cm, critical superconducting transition temperature above 8.9 K and residual resistance ratio over 6.4. We suggest a modified kinetic model to predict Nb films stress with the linear dependence of stress on the working pressure replaced with an exponential one, which allowed reduction of the average percent approximation error from 20 to 8%.

## 1. Introduction

The key elements of single quantum logic (SFQ) circuits are Nb/Al/$AlO_x$/Nb Josephson Junctions (JJ) and Nb-inductors, which are extremely promising for high-performance computing [1], [2], [3], [4] particularly as classical coprocessors for control and error tracking of large-scale quantum arrays [5], [6], [7]. In addition, Nb/Al/Al/$AlO_x$/Nb structures are the base of high-frequency receivers [8], parametric amplifiers [9], voltage standards [10] and highly sensitive magnetic field sensors [11]. The main material used in SFQ-circuits is thin-film niobium [12], [13], which can also be used [14] to increase the quality of superconducting circuits [15] and for the thermoelectric applications [16]. There are many studies on magnetron sputtering parameters for the growth of high-quality polycrystalline Nb films evaluating the roughness (Rq≈5 Å) [17], [18], intrinsic stress (from -600 to 600 MPa) and roughness (Rq≈7 Å) [19], [20], [21], residual resistance ratio (RRR) and intrinsic stress (RRR=3.9 and Rq≈24 Å) [22], critical superconducting transition temperature ($T_C$, from 8.0 to 9.1 K), and intrinsic stress (from -600 to 800 MPa) [23], [24], [25]. There is a work with a more detailed research [26], including a joint analysis of the dependence of intrinsic stress (from -500 to 400 MPa), $T_C$ (from 7.1 to 9.0 K) and critical current on the Nb films sputtering parameters. However, a comprehensive study of the mutual effect of magnetron sputtering parameters on the properties of niobium films on $SiO_2$ for SFQ-circuits at room and cryogenic temperatures is still uncovered. We believe it is necessary to investigate the properties of Nb thin films such as roughness, intrinsic stress, electrical resistivity, $T_C$ and RRR as part of a unified study, which will improve the reliability and performance of superconducting circuits.

Here we report a comprehensive study of the influence of magnetron sputtering parameters on the properties of niobium thin films on $SiO_2$ and Nb-based resistors for 1 and 10 kΩ nominal values at room and cryogenic temperatures. The relationship between the mechanical, electrical and cryogenic properties of Nb films is demonstrated. We show that there is a critical value of the working pressure $p_{critical}$ (3.0 mTorr or 0.4 Pa in our case) at a fixed sputtering power, above which stress in the film relaxes but its characteristics (roughness, conductivity, critical temperature, residual resistance ratio) decrease. We believe that this trend is universal and relevant for other sputtered metallic films such as Mo, W, Re, V, Hf and Rh. More than twenty 100-mm wafers with Nb films and 8 samples with Nb-based resistors are investigated, providing a wide range of experimentally measured 200-nm Nb-film properties. The results of the study provide a control of intrinsic stress in 200-nm Nb-film in the range from compression 400 MPa to tensile 600 MPa maintaining perfect parameters of the films: roughness Rq less than 0.8 nm, electrical resistivity less than 20 μΩ×cm, critical superconducting transition temperature more than 8.9±0.1 K and residual resistance ratio more than 6.4±0.2. A kinetic model of intrinsic stress generation is used to predict stress as a function of sputtering working pressure. An improvement of the kinetic model for high-energy methods of thin film deposition was proposed and allowed the average percent approximation error improvement from 20 to 8%.

## 2. Experiments

100-mm silicon wafers with thermal 3-μm-thick oxide on the surface are used as substrates for Nb thin films for room temperature characterization (≈ 305 K) of stress, electrical resistivity and roughness. Nb is deposited on 25×25 mm 3-μm $SiO_2$/Si-substrates to form Nb-based resistors (Nb meanders with normal state resistance of 1 and 10 kΩ at room temperature and 4 and 2 μm width) for cryogenic characterization of RRR and $T_C$. All of the deposition conditions are the same for different types of substrates. The wafers are not clamped in the holders, so they are deformed only by the intrinsic stress in the films. Substrates are cleaned with megasonic cleaning and Piranha solution (mixture of $H_2SO_4$ and $H_2O_2$) before deposition. Thin 200-nm Nb films are deposited in an ultrahigh vacuum magnetron sputtering system with 3″ (76.2-mm) targets and with the base pressure below $5\times10^{-7}$ Pa as in ref [27]. The working pressure of argon $p_{Ar}$ is varied from 0.3 to 7.0 mTorr (from 0.04 to 0.93 Pa) and the sputtering power $P_{sput}$ is varied from 250 to 400 W, while the distance between the sputtering target and the substrate is between 80 and 130 mm. Substrate heating for Nb deposition is not allowed for the SFQ circuits [12], [20], so we cannot use our approaches to form ultra-smooth films [28], [29]. All films in our experiments were deposited at 20°C. The processing of the Nb-based resistors on 25-mm substrate is done using laser lithography and $CF_4$ plasma-based ICP RIE etching. The chip for cryogenic characterization is cut into 10mm$^2$ dies prior to measurements. Design details are shown in the supplementary materials.

The electrical resistivity $\rho_{Nb}$ of Nb films is measured at 64 points on the entire 100-mm wafers using the 4-probe method. This parameter is important because, according to the Mayadas–Shatzkes' grain boundary scattering model, an increase in resistivity indicates structural defects in the film [30]. Low surface roughness is critical for the uniform tunneling current in the Nb/Al/AlOx/Nb JJ [31]. We measured Rq at 25 points across the entire wafer using a stylus profiler. Intrinsic stress in the films is measured, since high stress values result in a decrease of both Nb/Al/AlOx/Nb JJ' critical current reproducibility [19] and the quality factor of Josephson junctions [23]. The stress is calculated from the change in wafer curvature after the deposition of the Nb film. The initial and final wafer curvature is determined using a stylus profiler. High-resolution SEM images are acquired for all Nb films to measure the grain size.

For cryogenic characterization, the samples are cooled down using a pulse tube cryocooler at a base temperature of 2.5 K. For Nb-based resistors, the RRR and $T_C$ are measured with an accuracy of ±0.05 K. RRR and $T_C$ represent the quality of the thin film material structure (defects, grain boundaries). RRR is calculated as the ratio of the resistance of the Nb-based resistors at 300 K and 10 K measured with the four-probe method. A direct relationship between RRR and the energy relaxation time $T_1$ is demonstrated [32]. An inverse relationship between $T_C$ and the number of structural defects has been shown for quantum superconducting circuits with a Nb base layer [26].

## 3. Results

### *3.1 Surface and electrical film parameters as dependence on sputtering parameters*

The main magnetron sputtering parameters are the working pressure $p_{Ar}$ and sputtering power $P_{sput}$; therefore, we studied the influence of these parameters on the properties of Nb films. Figure 1,a shows the dependence of Nb film roughness on $p_{Ar}$ (each point on the plot corresponds to one Rq measurement and green and yellow points indicate two different wafers with identical films). At $p_{Ar}$ less than 3 mTorr (0.4 Pa), the average roughness Rq on a 100-mm wafer does not exceed 8 Å, but the higher $p_{Ar}$ results in a dramatic increase in Rq up to 18 Å. Figure 1,c shows the dependence of the resistivity of Nb films on $p_{Ar}$ at the fixed $P_{sput}$ of 400 W. It can be seen from figure that both the resistivity $\rho_{Nb}$ of the films and the variation (3σ) of $\rho_{Nb}$ values across the wafer increase at $p_{Ar}$ greater than 2.0 mTorr (0.27 Pa). The $P_{sput}$ in the range of 250 – 400 W had no noticeable effect on the $\rho_{Nb}$ and roughness of Nb films.

The effect of $P_{sput}$ at a fixed $p_{Ar}$ of 1.0 mTorr (0.13 Pa) on the stress in Nb films is shown in Figure 1,d. As $P_{sput}$ increases, the stress becomes more compressive, but the effect of $P_{sput}$ is small compared to the effect of $p_{Ar}$, as shown in Figure 1,e (each point on the plot corresponds to one 100 mm wafer). As $p_{Ar}$ increases from a minimum value of 0.3 mTorr (0.04 Pa), the stress changes from compressive to tensile. At a $p_{Ar}$ value around 3.0 mTorr (0.4 Pa), which we named $p_{critical}$, the stress begins to decrease from the maximum tensile value. Figure 1,f shows the stress map for the Nb film with stress values ranging from -170 to 70 MPa and a wafer average of 25 MPa. Wafer-to-wafer variability for average stress measurements is found to be less than 100 MPa.

A well-known issue of magnetron sputtering is a target erosion effect on the film properties, including film stress. There are a number of methods to control stress in Nb thin films which have been studying for many years. Imamura et al. [26] proposed cathode voltage as a universal magnetron sputtering parameter to control the stress. However, Booi et al. [33] showed that the stress-cathode voltage relationship changes during target erosion and proposed to control the ratio of cathode current $I_C$ and $p_{Ar}$ to maintain stress value constant. Then Amos et al. [34] showed that the stress for the $I_C/p_{Ar}$ ratio changes with target erosion and proposed to experimentally detect a single point on the $p_{Ar}$-$I_C$ plane that provides zero stress during the entire lifetime of the target. Modern magnetron power supplies allow a constant power mode, and it is reported in [35] that this mode provides stability of the stress during target erosion.

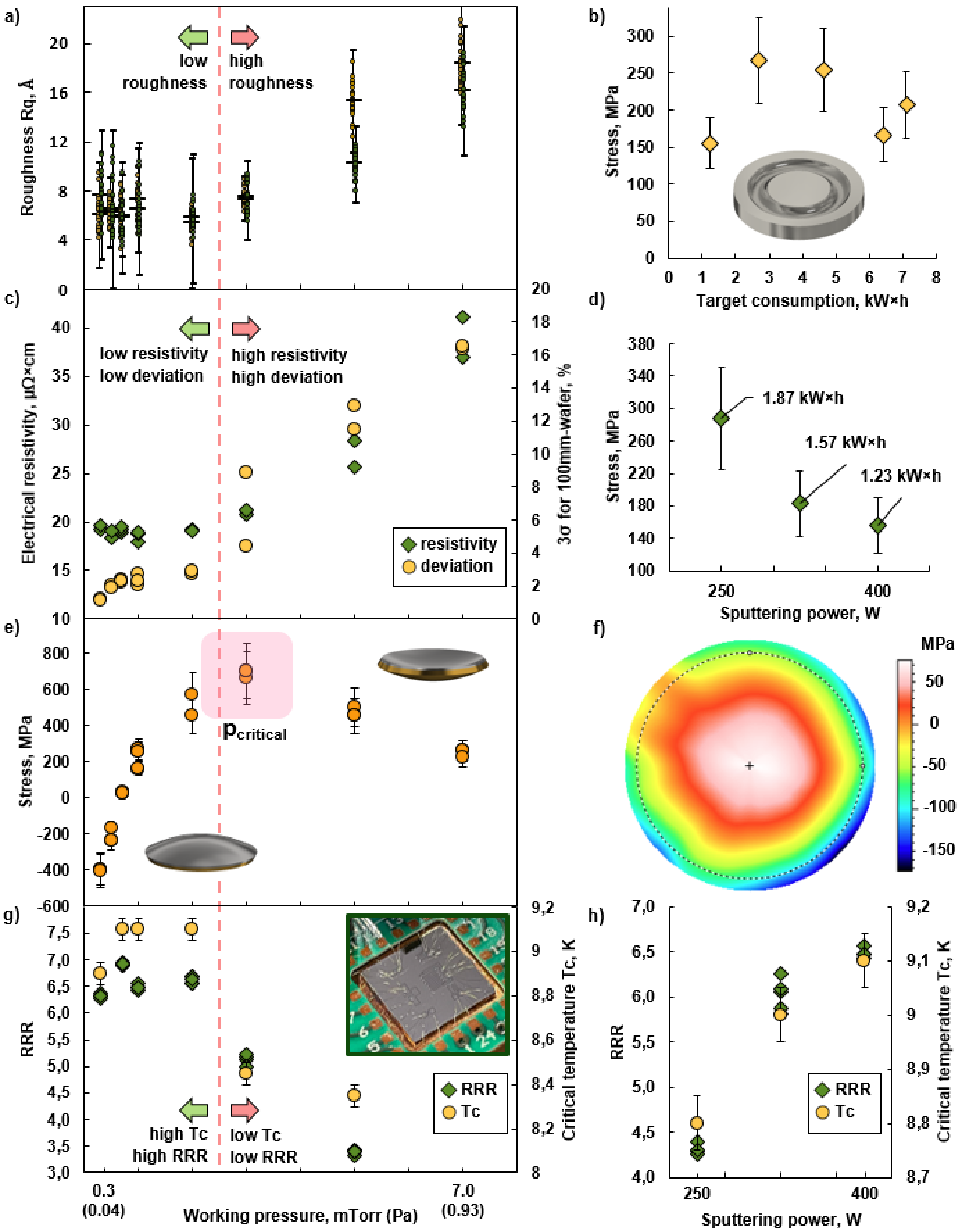

Figure 1. (a) Nb-films roughness Rq on a 100-mm wafer versus working pressure $p_{Ar}$ (green and yellow points next to each other indicate two different wafers with identical films). (b) Stress in Nb films as a function of magnetron target lifetime. (c) The electrical resistivity average value and scatter (3σ, %) on a 100-mm wafer of Nb films versus the $p_{Ar}$. (d) Stress in Nb-films as a function of sputtering power $P_{sput}$. (e) Stress in Nb-films as a function of the $p_{Ar}$. (f) Example of a map of stress distribution on a 100-mm wafer. (g) RRR and $T_C$ of Nb-based resistors versus the $p_{Ar}$ during the deposition of Nb films; the inset shows a photograph of a 10×10 mm chip for cryogenic characterization. (h) RRR and $T_C$ Nb-based resistors versus the $P_{sput}$ during the deposition of Nb films.

We measured the stress of our Nb films deposited at $P_{sput}$ 400 W and $p_{Ar}$ 1.0 mTorr (0.13 Pa) at target consumption from 1 to 7 kW×h (target resource 10 kW×h) and found a slight and non-monotonic change in the stress, as shown in Figure 1,b. Thus, the fixation of $P_{sput}$ and $p_{Ar}$ ensures the stability of the stress throughout the target lifetime.

### *3.2 $T_C$ and RRR film parameters as dependence on sputtering parameters*

The cryogenic properties of Nb films are investigated using a 4-probe method by measuring the voltage on Nb-based resistors with room temperature resistance of 1 and 10 kΩ. An electrical connection of the test chip in the holder is shown in the inset of Figure 1,g. The dependence of the RRR and $T_C$ of the Nb-based resistors on $p_{Ar}$ at a fixed $P_{sput}$ of 400 W is shown in Figure 1,g. Each point on the plot refers to one resistor on the chip. Chip-to-chip variability for RRR and $T_C$ measurements is found to be less than 0.2 and 0.1 K. $T_C$ and RRR decrease at $p_{Ar}$ greater than certain $p_{Ar}$. The dependence of RRR and $T_C$ of Nb-based resistors on $P_{sput}$ at a fixed $p_{Ar}$ of 1.0 mTorr (0.13 Pa) is shown in Figure 1,h. $T_C$ and RRR increase monotonically with increasing $P_{sput}$, which is consistent with the data of [22].

### *3.3 Kinetic model for stress prediction in Nb films*

Prediction of a thin films stress depending on the deposition parameters is an important problem as it plays a prominent role in the reliability, reproducibility and functionality of thin film devices [36], [37], [38]. To the date, the stress formation in thin films is best described by the kinetic model by Chason et al. [36], [39], in particular for high-energy methods like magnetron sputtering [40], [41]. It is capable predicting the change of intrinsic film stress as a function of its thickness. However, in our experiment the stress is measured *ex situ* in a fully-formed film. Therefore, it is assumed that the stress is uniform over the film thickness. In this case, the final expression for the full stress in the film is

$$\sigma_{ss}^{sputt} = \sigma_{growth} + \sigma_{gb}^{energetic} + \sigma_{bulk}^{energetic} = \left[\sigma_C + (\sigma_T - \sigma_C)e^{-\frac{\beta D_{eff}}{RL}}\right] + A_0\left(\frac{l_0}{L}\right) + \left(1 - \frac{l_0}{L}\right)\frac{B_0}{\left(1 + \frac{l_0}{R\tau_S}\right)}, \quad (1)$$

where $\sigma_{growth}$ is the stress resulting from the formation and densification of grain boundaries. $\sigma_{gb}^{energetic}$ is compressive stress resulting from the densification of grain boundaries due to exposure of high-energy particles. $\sigma_{bulk}^{energetic}$ is compressive stress resulting from the point defects generation in the grain bulk and their annihilation on the free surface due to exposure of high-energy particles. $\sigma_C$ is compressive stress created by the insertion of adatoms into grain boundaries under the effect of non-equilibrium chemical potential at the surface (fitting parameter). $\sigma_T$ is the tensile stress resulting from the formation of new segments of grain boundaries (fitting parameter). $\beta D_{eff}$ is an exponential term describing the efficiency of diffusion of adatoms from the free surface into the grain boundaries (fitting parameter). $R$ is the growth rate of the grain, i.e., the deposition rate of the film. $L$ is the average grain size. $l_0$ is the depth of defect generation by a high-energy particle in the grain bulk and the width of the grain boundary region where direct densification occurs under the effect of a high-energy particle (fitting parameter). $A_0$ is a fitting parameter that defines $\sigma_{gb}^{energetic}$. $B_0$ is a fitting parameter that defines $\sigma_{bulk}^{energetic}$ and directly proportional to the defect generation rate in the grain bulk under high-energy exposure. The next explicit dependence on the working pressure $p$ is introduced for the parameters $A_0$, $B_0$, and $l_0$:

$$A_0 = A^*\left(1 - \frac{p}{p_0}\right), \quad (2)$$

$$B_0 = B^*\left(1 - \frac{p}{p_0}\right), \quad (3)$$

$$l_0 = l^*\left(1 - \frac{p}{p_0}\right), \quad (4)$$

where $p_0$ is certain pressure value that is the same for all experiments in which only $p$ is varied. $A^*$, $B^*$, $l^*$ are fitting coefficients included in linear dependence of $A_0$, $B_0$, and $l_0$ on $p$. They reduce the total number of fitting parameters for a large number of experiments with different $p$.

$\tau_S$ is the characteristic time required for diffusion of a point defect from the grain bulk to the free surface and is determined by the expression:

$$\tau_S = \frac{l_0}{R}\left[\alpha - 1 - \alpha\sqrt{1 - \left(\frac{2}{\alpha}\right)}\right], \quad (5)$$

where $\alpha = \frac{D_i}{2Rl_0}$ for $\alpha > 2$ and $\alpha = 2$ for $\alpha \leq 2$, where $D_i$ is the defect diffusion parameter (fitting parameter). When $\alpha \leq 2$ all defects remain in the grain bulk.

The kinetic model can consider [42] the effect of the transverse grain size variation on the stress [43]. However, because our deposition regime is in zone I of the Thornton diagram [44], the transverse grain size can be considered constant over the entire grain height, similar to molybdenum films [45]. In the case of metals with low melting point, relaxation of the intrinsic stress occurs after the thin film deposition, but for refractory metals, the level of stress is maintained [46]. Therefore, we consider that the stress we measured is equal to the stress immediately after Nb-deposition.

We approximated our experimental data on $p_{Ar}$ - dependent stress using the non-linear least squares fitting routine. The values of stress (average), working pressure, grain size, and deposition rate used for fitting are presented in Table 1. The grain size $L$ is defined as the diameter of a circle with a length equal to the perimeter of the grain in the SEM image. Due to the extremely low contrast, the grain boundary is traced manually. Examples of manual contouring and grain size histograms are shown in the supplementary materials. The grain size dependence of the Nb film on $p_{Ar}$ is shown in Figure 2,a. SEM images of the Nb films deposited at $p_{Ar}$ 0.3, 0.7, 3.0, and 7.0 mTorr (0.04, 0.09, 0.40 and 0.93 Pa) are shown in Figures 2,b, 2,c, 2,d, and 2,e, respectively. Cross-section SEM images of these films are shown in the supplementary material. In Figure 2,f and Figure 2,g, the grain size $L$, nm is calculated as a linear function of the working pressure $p, mT$:

$$L = 2.7287 \times p + 47.95. \quad (6)$$

In model plots, the deposition rate $R, nm/s$ is calculated as a linear function of the working pressure $p, mT$:

$$R = -0.0006 \times p + 0.2898. \quad (7)$$

Table 1. Input experimental data for the kinetic model.

| № | Stress, GPa | $p_{Ar}$, Pa | L, nm | R, nm/s |
|---|---|---|---|---|
| 1 | -0.400 | 0.0399 | 48.1 | 0.290 |
| 2 | -0.203 | 0.0665 | 49.8 | 0.290 |
| 3 | 0.028 | 0.0931 | 49.4 | 0.290 |
| 4 | 0.211 | 0.1330 | 51.6 | 0.288 |
| 5 | 0.511 | 0.2660 | 52.5 | 0.288 |
| 6 | 0.681 | 0.3990 | 54.6 | 0.287 |
| 7 | 0.476 | 0.6650 | 66.4 | 0.287 |
| 8 | 0.242 | 0.9310 | 64.5 | 0.286 |

The published fitting parameter values for refractory materials are summarized in Table 2 along with the assumed value bounds and the starting values for the fitting parameters.

Table 2. Values of fitting parameters of the kinetic model from the literature and their fitting starting values and bounds.

| ref. | film | dep. | $\sigma_t$, GPa | $\sigma_c$, GPa | $P_0$, Pa | $\beta D_{eff}$, nm²/s | $D_i$, nm²/s | $A^*$, GPa | $B^*$, GPa | $l^*$, nm |
|---|---|---|---|---|---|---|---|---|---|---|
| [40] | Mo | DC | 4.24 | -1.6 | 0.83 | 0.28 | 7.2 | -167 | -53.1 | 1.7 |
| [40] | Mo | HiPIMS | 4.24 | -1.6 | 0.83 | 1.29 | 7.2 | -115.2 | -243 | 1.8 |
| [41] | Mo | DC | 0.78 | -29.9 | 0.67 | 0.03 | 0.68 | -2.60 | -12 | 0.7 |
| [47] | W | DC | 3.16 | -4 | 0.52 | 6e-9 | 0.24 | -8.13 | -11.57 | 0.62 |
| [48] | Ti | Evap | 1.08 | -0.16 | - | 1.87 | - | - | - | - |
| Start | Nb | DC | 2 | -20 | 0.6 | 0.03 | 0.6 | -3 | -12 | 0.8 |
| Min | Nb | DC | 0.5 | -40 | 0.13 | 0 | 0 | -10 | -15 | 0.5 |
| Max | Nb | DC | 8 | -1 | 1 | 1 | 10 | 0 | 0 | 4 |

Figure 2,f demonstrates the model fitting for the 8 experimental points (orange line). The model predicts an extremum in stress, as well as the experimental data. However, the model predicts that the stress will become increasingly compressive with further $p_{Ar}$ increasing, but it was shown before [19], [26], that it should relax to zero. If the first 6 points only are used for model fitting (without 5.0 and 7.0 mTorr points), the model also predicts a compression stress transition (orange line in Figure 2,g). The reason for this is the linear dependence of the parameters $A_0$, $B_0$, and $l_0$ on the working pressure.

In this case, these parameters, corresponding to the effect of high-energy exposure on the stress in the film, are the same over the entire range of $p$. However, it is reasonable to assume that when some $p$ is exceeded, the excess energy of the deposited atoms and "fast neutrals" is dissipated before reaching the substrate. Therefore, we suggest replacing the linear dependence of the parameters $A_0$, $B_0$, and $l_0$ on $p$ with an exponential ones:

$$A_0 = A^* \times p^{*\left(\frac{p}{p_0}\right)}, \quad (8)$$

$$B_0 = B^* \times p^{*\left(\frac{p}{p_0}\right)}, \quad (9)$$

$$l_0 = l^* \times p^{*\left(\frac{p}{p_0}\right)}, \quad (10)$$

where $p^*$ is a dimensionless exponential fitting parameter less than 1.0. In this case the effect of the parameters $A_0$, $B_0$, and $l_0$ on the stress decreases as $p$ increases and asymptotically tends to zero. Figure 2, g shows in green the result of the model fitting using 6 experimental points with exponential coefficients. One can see that such a model does not predict unexplained changes in the stress value. For all the 8 experimental points, the result of model fitting with exponential coefficients is shown in Figure 2,f (green line). The results for the fitting parameters in the case of linear/exponential coefficients and for 6/8 experimental points are given in Table 3.

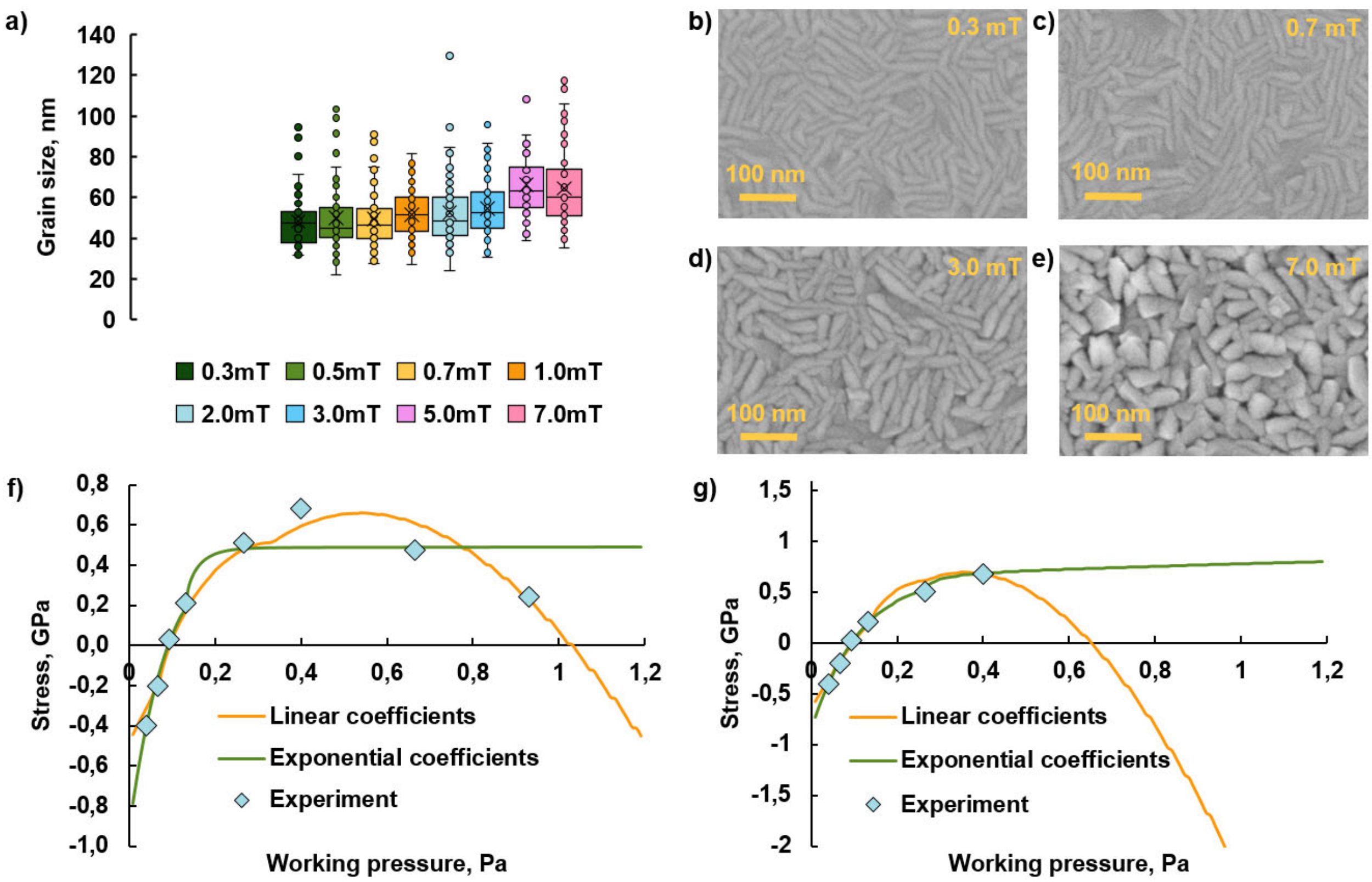

Figure 2. (a) Grain size dependence of Nb films on working pressure $p_{Ar}$. SEM images of Nb films deposited at $p_{Ar}$ (b) 0.3 mTorr (0.04 Pa), (c) 0.7 mTorr (0.09 Pa), (d) 3.0 mTorr (0.4 Pa), (e) 7.0 mTorr (0.93 Pa). Results of the kinetic model approximation of the experimental dependence of stress in Nb films on the $p_{Ar}$ for linear (orange line) and exponential (green line) coefficients for 8 points (f) and 6 points (g).

Table 3 - Values of the fitting parameters of the kinetic model for the cases of linear and exponential coefficients and 6 and 8 experimental points. The final parameters are shown in color.

| Coef. | $\sigma_t$, GPa | $\sigma_c$, GPa | $P_0$, Pa | $\beta D_{eff}$, nm$^2$/s | $D_i$, nm$^2$/s | $A^*$, GPa | $B^*$, GPa | $l^*$, nm | $p^*$ |
|---|---|---|---|---|---|---|---|---|---|
| lin. 8 | 0.62 | -18.67 | 0.33 | 0.08 | 3.19 | -7.34 | -0.96 | 3.55 | - |
| exp. 8 | 0.50 | -1.00 | 0.42 | 0.12 | 2.55 | -7.72 | -1.47 | 4.00 | 0.06 |
| lin. 6 | 1.08 | -22.22 | 0.27 | 0.31 | 2.15 | -9.83 | -0.87 | 3.99 | - |
| exp. 6 | 1.19 | -11.80 | 0.38 | 0.63 | 0.28 | -2.06 | -2.88 | 2.59 | 0.03 |

The suggested kinetic model with exponential coefficients allows improving the average percent approximation error from 20 to 8%, calculated using the formula:

$$\delta = \frac{1}{n}\sum_{i=1}^{n}\left|\frac{\sigma_i - \sigma_{approx\ i}}{\sigma_i}\right| \times 100\%, \qquad (1$$

where $n$ is the number of points, $\sigma$ is the experimental value of stress, $\sigma_{approx}$ is the calculated value of stress.

All calculated fitting parameters have adequate values. The value of the kinetic (independent of high-energy effects) diffusion parameter of adatoms $\beta D_{eff}$ for Nb is 0.63 nm$^2$/s, which is higher than the value for Mo 0.28 nm$^2$/s [40], but lower than one for Ti 1.87 nm$^2$/s [48]. This corresponds to the assumption [41] that the diffusion parameter is determined by the material and depends inversely on the melting temperature (2750 K for Nb, 2896 K for Mo and 1941 K for Ti). The value of the defect diffusion parameter (high-energy parameter) $D_i$ is 0.28 and it is much closer to the value for W 0.24 nm$^2$/s [47] than for Mo 0.68 nm$^2$/s [41], although this parameter is also determined by the melting temperature (3695 K for W). This may be due to the exclusion of experimental points above $p_{critacal}$ in our work. The adequacy of the $A^*$ and $B^*$ values is difficult to assess at this stage as the published values for Mo vary widely from -2.6 [41] to -167.0 [40] and -12.0 to -53.1. The defect generation depth $l^*$ for Nb 2.59 nm exceeds the values for Mo 0.70 nm [41] and W 0.62 mm [47], since in our experiment the distance between the sputtering target and the substrate is between 80 and 130 mm, while in the literature data it was 160 - 180 mm [41]. The kinetic parameters $\sigma_C$ and $\sigma_T$ have values close to those for other refractory metals like Mo and W, but it is important to compare them with the parameter values for Nb films obtained by a low energy method like electron beam evaporation. Unfortunately, these data are not available in the literature. The values of the parameters $p^*$ and $p_0$ are obtained for the first time for exponential function, so it is not possible to compare them with any published results. In this case $p_0$ is 2.8 mTorr (0.37 Pa) turned out to be in the range of $p_{critacal}$ about 3 mTorr (0.4 Pa), but theoretically $p_0$ can be both smaller and larger than $p_{critacal}$.

## 4. Discussion

We have shown that the key parameter of magnetron sputtering in niobium thin films deposition is the argon working pressure ($p_{Ar}$). The dependence of film roughness

Rq on $p_{Ar}$ confirms the published data [17], [19], [20], which indicate that Rq changes slightly with increasing $p_{Ar}$ up to a certain value of $p_{Ar}$. In this study, the roughness degraded dramatically at $p_{Ar}$ greater than 3 mTorr (0.4 Pa). The resistivity of Nb at low $p_{Ar}$ is also almost constant, but at pressures above 2.0 mTorr (0.27 Pa) it begins to increase. It is important to note that the scatter of resistivity values across the wafer also begins to increase at $p_{Ar}$ greater than 2.0 mTorr. The roughness and resistivity data clearly indicate a degraded Nb film structure at $p_{Ar}$ above 3.0 mTorr, which corresponds exactly to $p_{critical}$. It is well known that sputtered films contain Ar atoms due to a shot peening effect and the Ar content increases abruptly when $p_{Ar}$ exceed certain thresholds. These threshold pressures coincide with the pressures where the tensile stress is maximum, i.e. $p_{critical}$ [26], [49]. Ar content increasing causes the porous microstructure observed by cross-sectional SEM [50]. As a result, the Rq and resistivity increase. These results may be important for other metals with low mobility for different applications, such as molybdenum [51], [52].

The dependence of stress in Nb films on $p_{Ar}$ provides important information. This type of dependence, in whole or in part, is often found in scientific publications [19], [20], [25], [26]. However, it has not been demonstrated together with the dependence of resistivity on $p_{Ar}$. If we compare these two behaviors, it is obvious that when $p_{critical}$ is exceeded the stress relaxes and the resistivity increases. According to the Mayadas–Shatzkes' grain boundary scattering model, an increase in resistivity means an increase the grain boundary reflection coefficient and indicates structural defects in the film [30], [53]. A similar dependence of the stress, structure, and morphology of the Nb film on the cathodic voltage during magnetron sputtering is shown in [26]. Thus, zero stress can be obtained in the range of small $p_{Ar}$ and large $p_{Ar}$, but in the second case, the structure and morphology of the film degraded (like dendritic-like structure in work [54]), and this degradation begins along with the relaxation of stress in the film.

The sputtering power in the range from 250– 400 W had no noticeable effect on the roughness and resistivity. This means that the critical power value below which the film structure degrades (porous microstructure and high roughness is more than 1.0 nm) [26] did not fall within the investigated $P_{sput}$ range.

Cryogenic characterization of Nb-based resistors shows that RRR and $T_C$ increase with increasing $P_{sput}$, which is in agreement with literature data [22], [26] and may be associated with a decrease in the ω-phase content in the film [55]. The dependence of RRR and $T_C$ on $p_{Ar}$ is very indicative, where for $p_{Ar}$ 0.3, 0.7, 1.0 and 2.0 mTorr (0.04, 0.09, 0.13 and 0.27 Pa), RRR and $T_C$ are almost the same. However, as $p_{Ar}$ surpasses $p_{critical}$, the values of RRR and $T_C$ decrease. The degraded cryogenic parameters of the Nb-based resistors also confirm the degraded structure of the Nb films deposited at $p_{Ar}$ above $p_{critical}$. The RRR value quantifies the overall level of impurities in niobium including: oxygen, nitrogen, hydrogen and argon [56], [57], so a decrease in RRR confirms an increase in the argon content in the film. It is worth noting that the influence of stress in Nb films on the parameters of Nb films and Nb-based resistors is slight if $p_{Ar}$ is below $p_{critical}$, i.e., $p_{Ar}$ is on the left branch of the working pressure-stress graph. We assume that when the $p_{critical}$ is exceeded, the film is not actually continuous, since internal stress relaxes due to plastic deformations and argon content in a film increases [50]. The post-growth incorporation of impurities due open grain-boundary structure also occurs. As a result, all film parameters degraded.

The introduction of exponential coefficients to determine $A_0$, $B_0$, and $l_0$ into the kinetic model does not allow us to describe the relaxation of stress after exceeding $p_{critical}$ value. However, our characterization data on Nb films and Nb-based resistors show that films at high $p_{Ar}$ contain structural defects and cannot be considered continuous, even if they appear so in SEM images. In other words, the kinetic model is not applicable to films formed at $p_{Ar}$ values above $p_{critical}$. For $p_{Ar}$ less than $p_{critical}$, it is more physical to use exponential coefficients that reduce the high-energy terms of the kinetic model asymptotically to zero.

## 5. Conclusion

We have performed a comprehensive study of the influence of magnetron sputtering parameters on the niobium films and Nb-based resistors properties on $SiO_2$ at room and cryogenic temperatures. We provided the experimentally obtained relationship between the mechanical, electrical, and cryogenic characteristics of the Nb films. We have shown that for Nb films, the key parameter of magnetron sputtering is the working pressure. We found critical working pressure $p_{critical}$ above which (3.0 mTorr or 0.4 Pa in our case) the intrinsic tensile stress starts to relax, but the roughness and resistivity increase, whereas the residual resistance ratio RRR and the superconducting transition temperature $T_C$ decrease. Below $p_{critical}$ it is possible to control change of intrinsic stress in 200-nm Nb-film in the range from compression 400 MPa to tensile 600 MPa while maintaining perfect for SFQ-circuits parameters of the films: roughness Rq less than 0.8 nm, electrical resistivity less than 20 μΩ×cm, $T_C$ more than 8.9 K and RRR more than 6.4. We have shown that the stress level in Nb films under the sputtering mode with fixed sputtering power $P_{sput}$ remains constant throughout the target lifetime. We found that maximizing $P_{sput}$ is necessary to increase RRR, and $T_C$. Finally, we used a kinetic model to predict the stress in Nb films and reduced the percent approximation error from 20 to 8% by replacing the linear dependence of the high-energy terms on the working pressure with an exponential one in it.

## Declaration of Competing Interest

The authors declare that they have no known competing financial interests or personal relationships that could have appeared to influence the work reported in this paper.

## Acknowledgements

Technology was developed and samples were fabricated at the BMSTU Nanofabrication Facility (Functional Micro/Nanosystems, FMNS REC, ID 74300).

## Author contributions statement

**EVZ:** Conceptualization (equal); Formal analysis (lead); Methodology (lead); Investigation (lead); Writing – original draft (lead); Visualization (lead).

**IAS:** Methodology (equal); Investigation (equal); Writing – review and editing (equal), visualization (supporting).

**SVB:** Methodology (equal); Investigation (equal).

**DAB:** Methodology (equal); Investigation (equal).

**MIT:** Methodology (equal); Investigation (equal).

**EAK:** Investigation (supporting); Writing – review and editing (equal).

**NSS:** Investigation (supporting); Writing – review and editing (equal).

**IAR:** Methodology (supporting); Writing – review and editing (supporting).

**SPB:** Formal analysis (supporting); Investigation (supporting).

**SAK:** Formal analysis (supporting); Investigation (supporting).

**NDK:** Methodology (supporting); Investigation (supporting).

**JAA:** Methodology (supporting); Investigation (supporting).

**IR:** Conceptualization (lead); Writing – review and editing (lead); Supervision (lead).

# Supplement to: Mutual control of critical temperature, residual resistance ratio, stress, and roughness for sputtered Nb films

E.V. Zikiy,[1,2] I.A. Stepanov,[1] S.V. Bukatin,[1] D.A. Baklykov,[1] M.I. Teleganov,[1] E.A. Krivko,[1] N.S. Smirnov,[1] I.A. Ryzhikov,[1,3] S.P. Bychkov,[1] S.A. Kotenkov,[1] N.D. Korshakov,[1] J.A. Agafonova[1] and I.A. Rodionov[1,2,*]

[1]Shukhov Labs, Quantum Park, Bauman Moscow State Technical University, Moscow, 105005, Russia

[2]Dukhov Automatics Research Institute (VNIIA), Moscow 127055, Russia
[3]Institute for Theoretical and Applied Electromagnetics RAS, Moscow 125412, Russia
*email: irodionov@bmstu.ru

This supplement provides experimental details to support the claims made in the main text.

### Scanning electron microscopy

Scanning electron microscopy was employed for grain size, thickness and uniformity assessment of deposited films. We used accelerating voltage 3-5 keV and probe current around 100 pA to observe samples with BSE/SE2 detector at working distance 2-3 mm. For assessing of the thin films thickness in cross-section 10 nm layer of Cr was deposited before SEM.

### Structure of thin films

Cross-section SEM images of thin films obtained by splitting are not a reliable source of information about the film structure, because the film is stretched and destroyed during splitting. However, some estimates can be made. In Figure S1,a, S1,b, S1,c, S1,d, we show cross-section SEM images of test films deposited at working pressures of 0.3, 0.7, 3.0, and 7.0 mTorr. It can be seen that the films obtained at 3.0 and 7.0 mTorr have a much more ‘loose’ structure. In addition, these films did not elongate like the films at 0.3 and 0.7 mTorr, which indicates a weak mechanical bond between the grains.

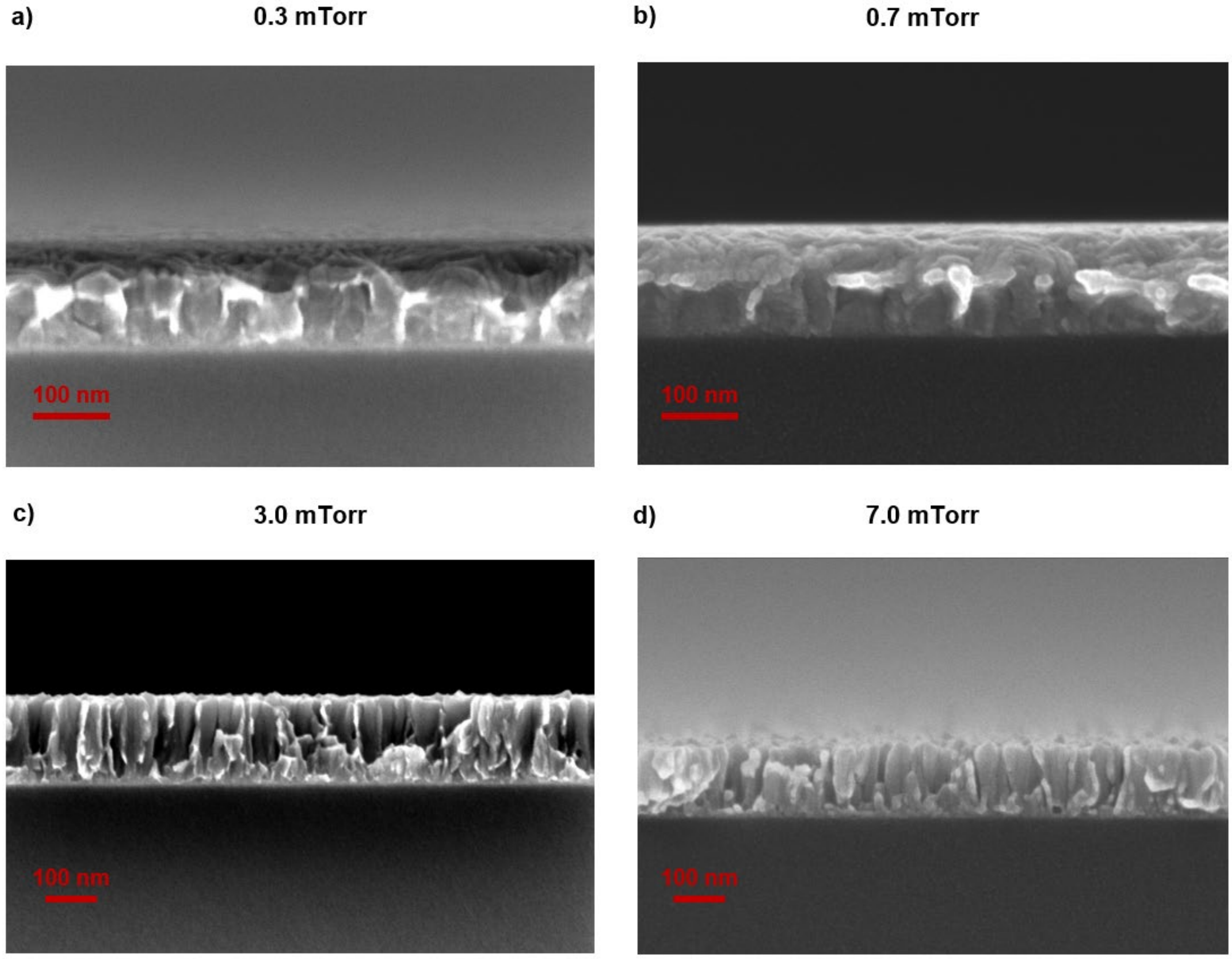

Figure S1. Cross-section SEM images of Nb films deposited at working pressure of 0.3 mTorr (a), 0.7 mTorr (b), 3.0 mTorr (c), 7.0 mTorr (d).

### Determination of grain size

The grain size $L$ is defined as the diameter of a circle with a length equal to the perimeter of the grain in the SEM image. Due to the extremely low contrast, automatic grain boundary detection (e.g., using ImageJ or MIPAR) is difficult, so the grain boundary is traced manually. Moreover, even manually, most of the grains cannot be contoured because it is impossible to differentiate the boundaries of neighboring grains. For each film, 64 grain sizes were obtained. Figures S1, S2, S3 and S4 show grain size distribution histograms and examples of manual contouring on SEM images for Nb films deposited at working pressures of 0.3, 0.7, 3.0 and 7.0 mTorr. It can be seen that the accuracy of this method of grain size determination is low, but for our purpose more accurate measurements (for example, using an atomic force microscope) are not required.

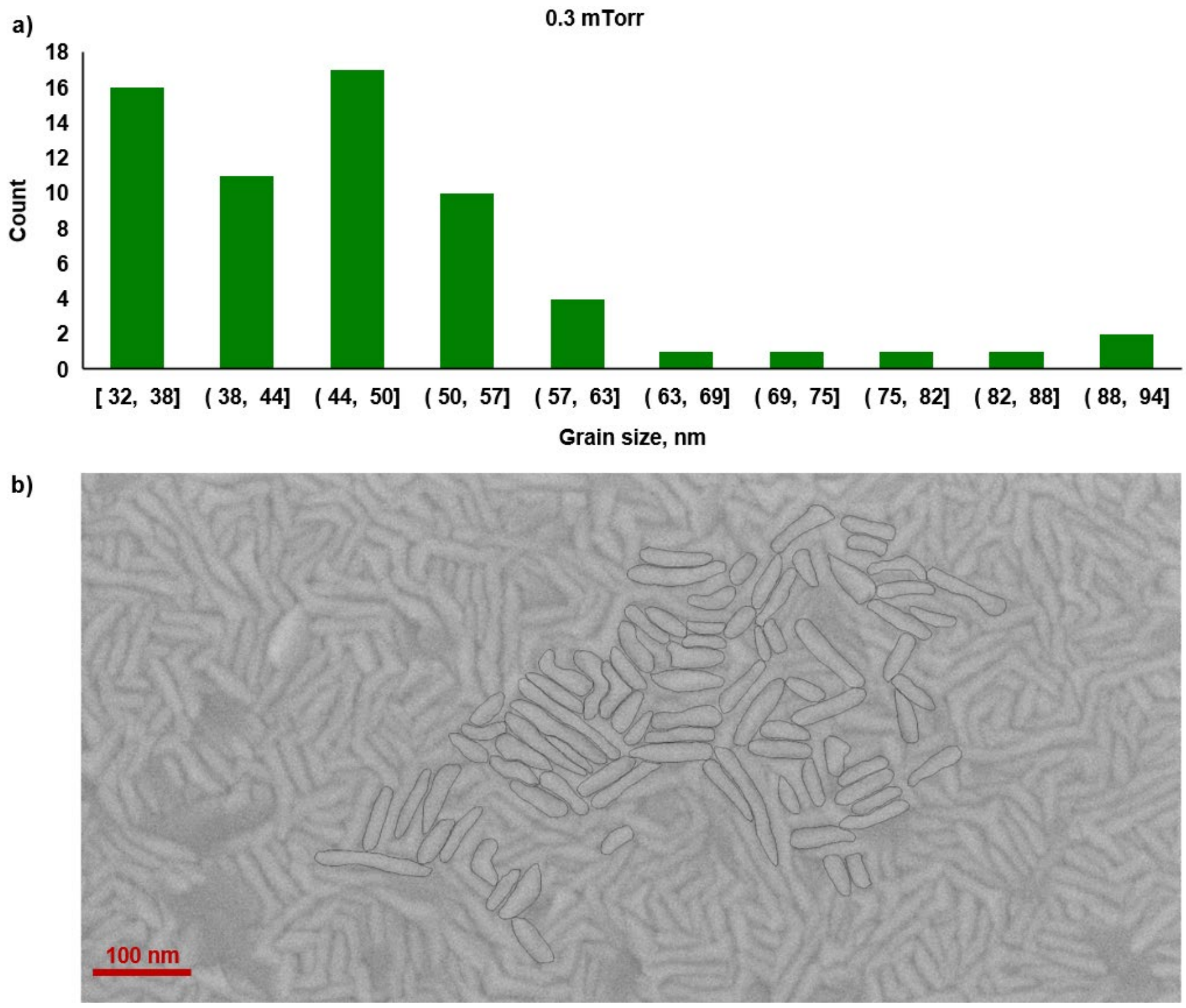

Figure S2. For Nb film deposited at $p_{Ar}$ 0.3 mTorr: (a) grain size distribution histograms and (b) high-resolution SEM image of grains.

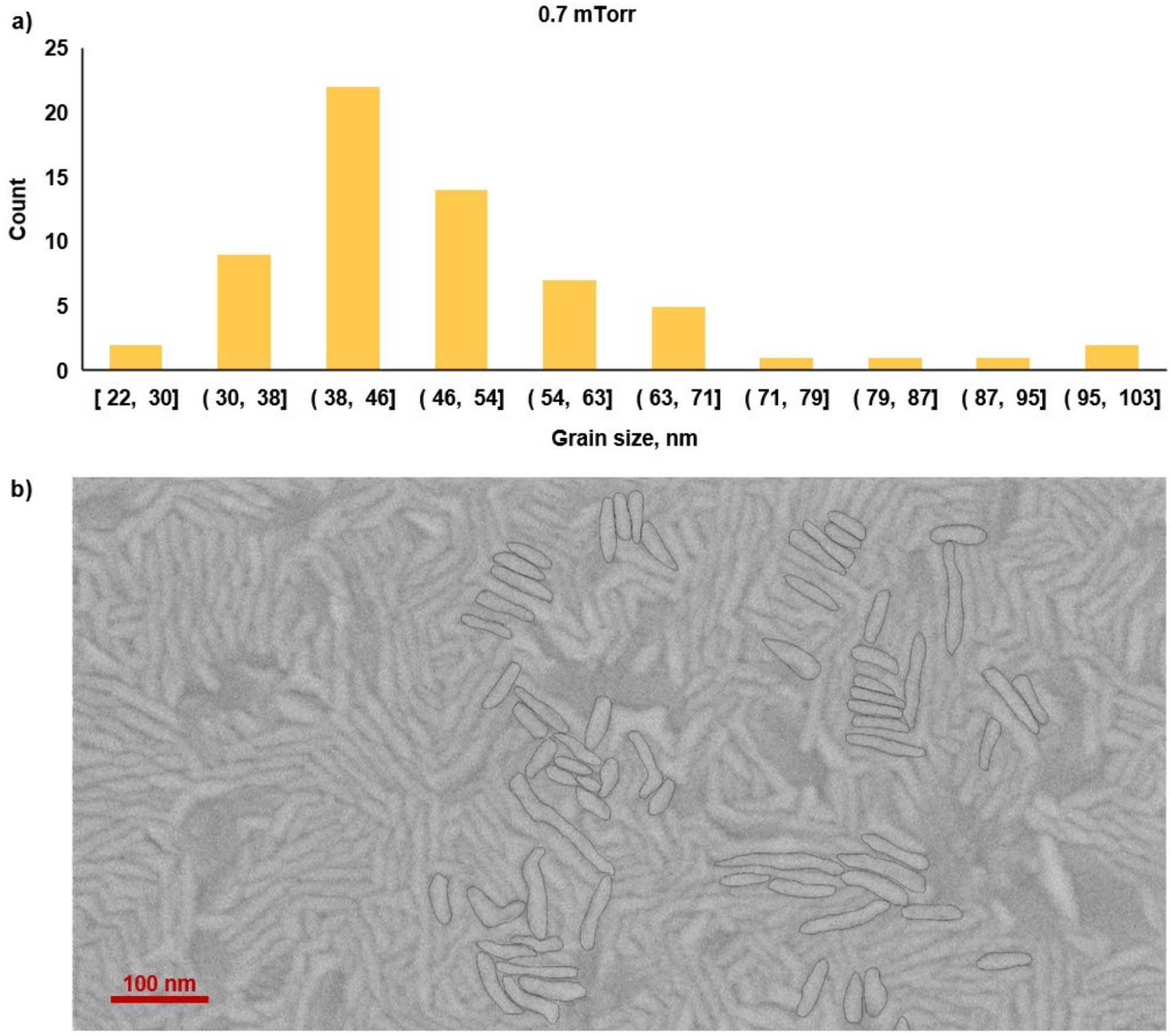

Figure S3. For Nb film deposited at $p_{Ar}$ 0.7 mTorr: (a) grain size distribution histograms and (b) high-resolution SEM image of grains.

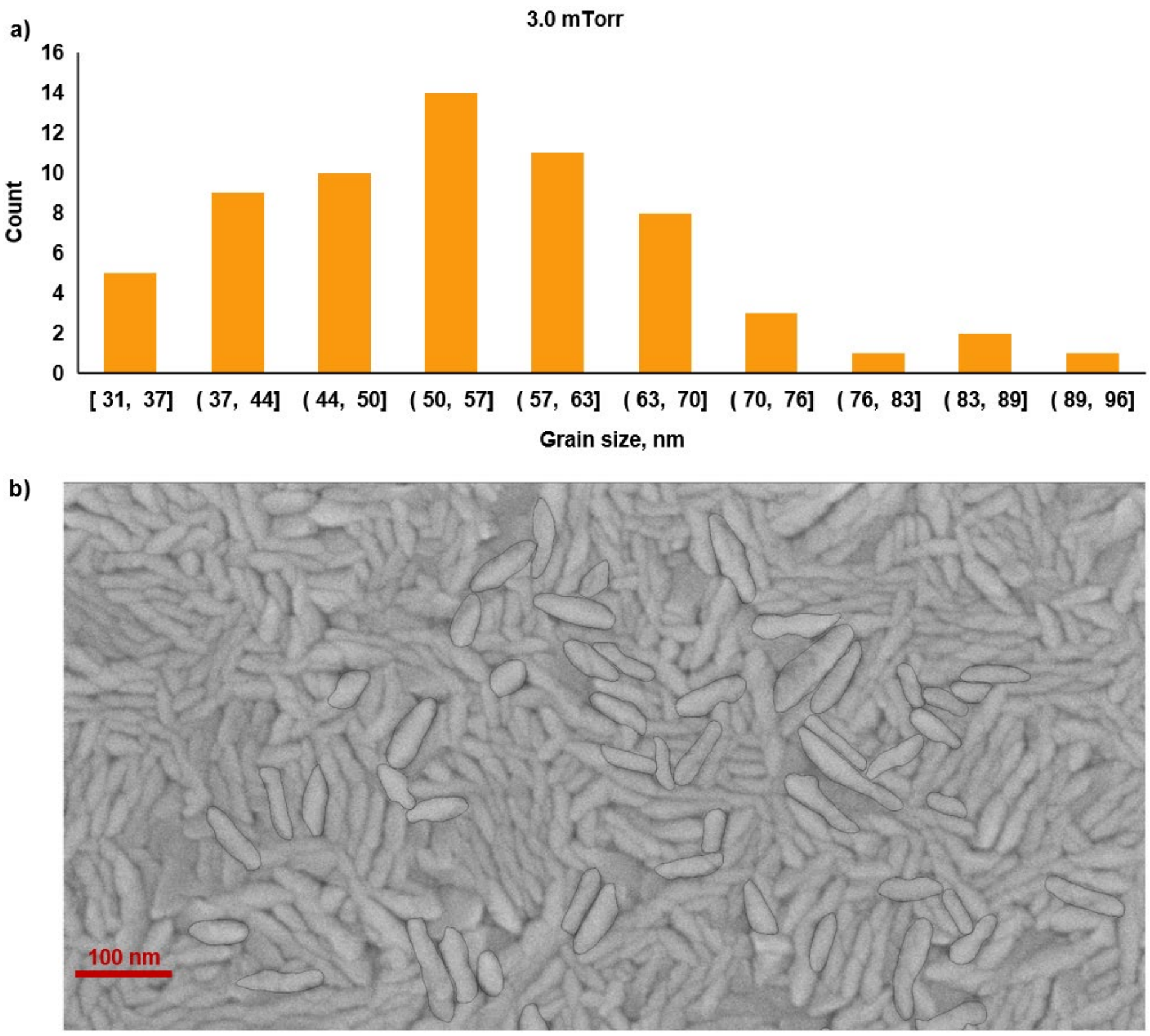

Figure S4. For Nb film deposited at $p_{Ar}$ 3.0 mTorr: (a) grain size distribution histograms and (b) high-resolution SEM image of grains.

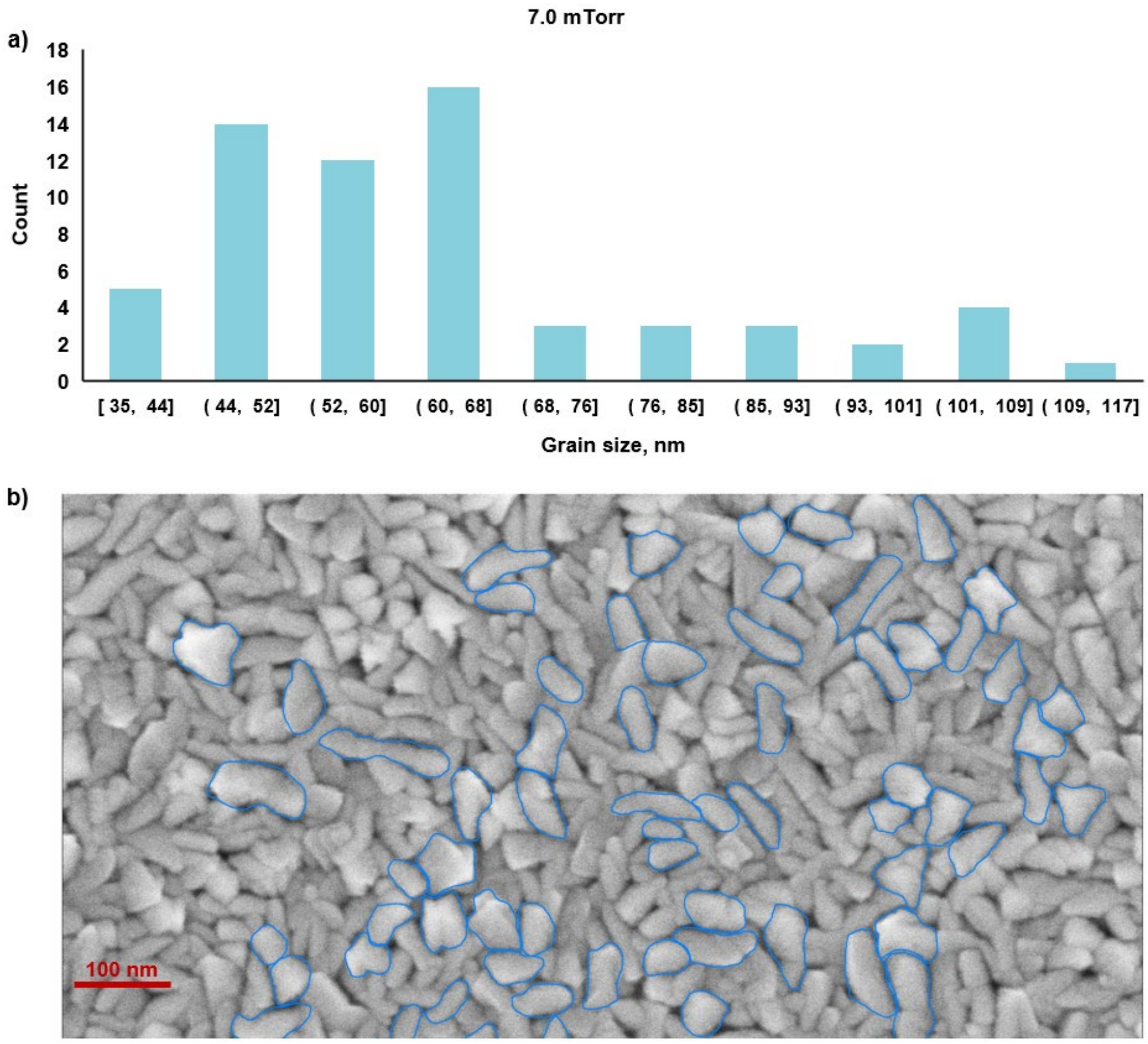

Figure S5. For Nb film deposited at $p_{Ar}$ 7.0 mTorr: (a) grain size distribution histograms and (b) high-resolution SEM image of grains.

### Design details of Nb-based resistors

The fabrication process of Nb-based resistors is shown schematically in Figure S6,a. The structures are formed using a subtractive process that includes laser lithography and plasma-chemical etching. Optical image of the Nb-based resistors with 1 and 10 kΩ resistance are shown in Figure S6,b and S6,c. Cross-section SEM image (at an angle) of Nb-meander is shown in Figure S6,d. Etching profile is shown in Figure S6,e. The etching regime is chosen with guaranteed etch into the substrate to prevent under-etching in case of etch rate decreasing upon changing the Nb-film deposition regime.

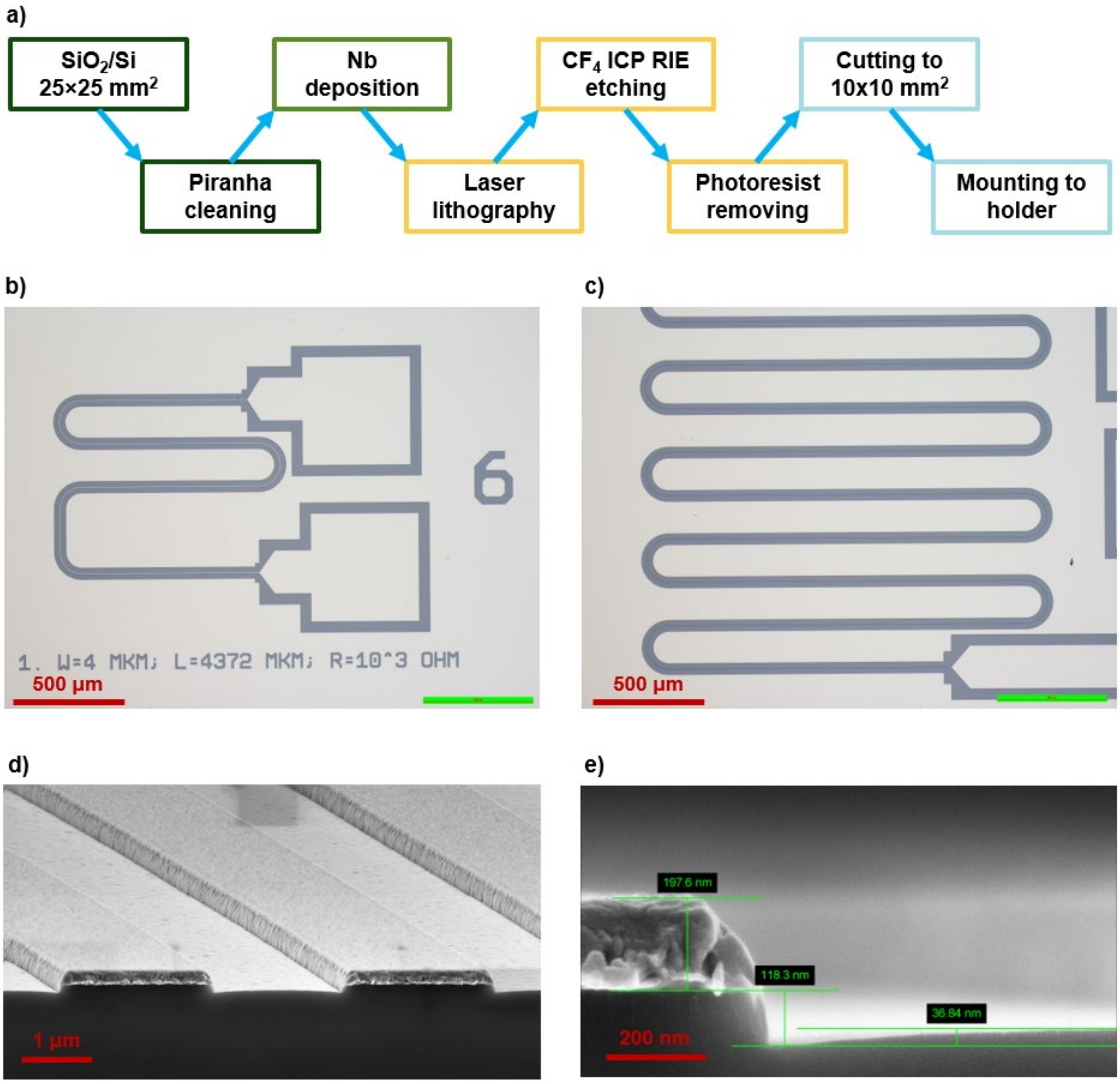

Figure S6. (a) Fabrication process. Optical image of the Nb-based resistors: (b) with line width 4 µm and resistance 1 kΩ and (c) with line width 2 µm and resistance 10 kΩ. SEM image of the Nb-based resistors cross-section: (d) at an angle and (e) etching profile at high magnification.